# Anomalous specific heat behaviour in the quadrupolar Kondo system $PrV_2Al_{20}$

**M Tsujimoto**[1] **Y Matsumoto**[1] **and S Nakatsuji**[1,2]

[1]Institute for Solid State Physics, University of Tokyo, Kashiwa, Chiba, Japan
[2]PRESTO, Japan Science and Technology Agency (JST), 4-1-8 Honcho Kawaguchi, Saitama 332-0012, Japan

E-mail: tsujimoto@issp.u-tokyo.ac.jp

**Abstract**. We have measured the specific heat of $PrV_2Al_{20}$ at very low temperatures, using high quality single crystals with the residual resistivity ratio ~ 20. The high-quality single crystals exhibit clear double transitions at $T_Q$ = 0.75 K and $T^*$ = 0.65 K. These transitions are clearer and shift to higher temperature in higher quality single crystals. Besides, there was no hysteresis in those transitions in warming and cooling process of the heat capacity measurements. In the ordered state below $T^*$, the specific heat shows a $T^4$ dependence, indicating the gapless mode associated with the quadrupole and/or octupole ordering.

## 1. Introduction

Hybridization between the conduction and $f$ electrons ($c$-$f$ hybridization) leads to a variety of interesting phenomena through the Kondo effect and its competition with the RKKY interaction, such as heavy fermion behavior, non-Fermi liquid behavior and unconventional superconductivity in the vicinity of quantum critical point (QCP) [1,2]. So far, most of the associated studies have been made for Ce ($f^1$) or Yb ($f^{13}$) based heavy fermion materials with significant magnetic and/or valence fluctuations [1-5]. A further interesting possibility is the QCP on the border of electric quadrupole order where the fluctuations of orbital degrees of freedom dominate. This is highly nontrivial since the ground state in the strong coupling limit between quadrupole and $c$- electrons itself is expected to be non-Fermi liquid due to the so called quadrupolar Kondo effect [6].

For the study of such a novel quantum criticality, the systems with cubic $\Gamma_3$ ground state doublet, realized in the cubic crystalline electric field (CEF) of a $f^2$ configuration, are supposed to be the best candidate where the ground doublet does not have the magnetic dipole. The possibility of the quadrupolar Kondo effect has been investigated extensively in the past few decades in several Pr or U based systems, such as $PrPb_3$ [7-9], $PrInAg_2$ [10] and $Y_{1-x}U_xPd_3$ [11]. However, so far, there has been no established example having cubic $\Gamma_3$ ground doublet with strong hybridization without structural disorders. On the other hand, recent studies suggested that $PrTr_2Al_{20}$ ($Tr$ =Ti, V) are the good candidates for the study of the quadrupolar Kondo effect, where the hybridization is supposed to be strong [12]. The crystal structure of $PrTr_2Al_{20}$ is the cubic $CeCr_2Al_{20}$-type with the space group Fd-3m [13], and these systems have cubic $\Gamma_3$ CEF ground state [12,14,15]. The long-range quadrupole ordering was observed at $T_Q$ = 2.0 and 0.6 K for $PrTi_2Al_{20}$ and $PrV_2Al_{20}$, respectively. While ferro-quadrupolar ordering has been confirmed for $PrTi_2Al_{20}$ by neutron scattering and ultrasonic measurements [14,15], antiferro-quadrupolar ordering has been suggested for $PrV_2Al_{20}$ from the robustness of $T_Q$ under magnetic field [12] and the existence of the high magnetic field phase [16].

There are several experimental evidence for the strong hybridization in these systems, such as -ln $T$ dependence of the electrical resistivity (the magnetic Kondo effect) at higher temperatures than the first excitation state of CEF ( 60 K (Ti) and 40 K (V)), the large Weiss temperatures (-55 K (V) and -40 K (Ti)), the observation of the Kondo resonance peak revealed by the resonant photoemission spectroscopy measurements for $PrTi_2Al_{20}$ [17], and the large hyperfine coupling constants revealed by NMR measurements [18]. Here, the hybridization is expected to be larger in $PrV_2Al_{20}$ compared to $PrTi_2Al_{20}$ due to the chemical pressure effect. The lattice parameter of $PrV_2Al_{20}$ (a=14.591(2) Å) is smaller than the one of $PrTi_2Al_{20}$ (a=14.723(7) Å) [13]. Interestingly, in $PrV_2Al_{20}$, anomalous temperature dependences have been observed above $T_Q$ in the electrical resistivity ($\rho \sim T^{1/2}$), magnetic susceptibility ($\chi \sim -T^{1/2}$) and specific heat ($C/T \sim T^{-3/2}$) [12]. These should be the strong hybridization effects and possibly arising from the quadrupolar Kondo effect, which is an interesting open question at this moment. Remarkably, heavy fermion superconductivity was observed recently in the both systems under ambient pressure at $T_c$ = 0.2 K (Ti) and 0.05 K (V), respectively [19,20]. The quasiparticle effective masses were estimated to be $16m_0$ (Ti) and $140m_0$ (V), showing larger effective mass for V system. Interestingly, in $PrTi_2Al_{20}$, $T_c$ and the effective mass become highly enhanced by applying pressure up to 1.1 K and $106m_0$ at 8.7 GPa where $T_Q$ starts decreasing, suggesting a putative QCP of quadrupole ordering [21].

Here, we present the results of the specific heat measurements for the pure single crystal of $PrV_2Al_{20}$ with RRR ~ 20. We note that the sample quality is very important for the study of the nonmagnetic Kondo effect based on the quadrupole degree of freedom because the degeneracy of non-Kramers cubic $\Gamma_3$ doublet can easily be removed by crystal defects. Indeed, in $PrV_2Al_{20}$, it has been already revealed that the quadrupole ordering only appears in relatively high quality samples with the residual resistivity ratio (RRR) larger than 4 [12,20,22]. We found the double peak structure in the temperature dependence of the specific heat at around the quadrupole ordering temperature, which is consistent with the observation made in the previous report using the sample with RRR ~ 7 [23]. Moreover, the double peak structure is more sharply observed in this work using the sample with RRR ~ 20, indicating that the double transition is intrinsic to $PrV_2Al_{20}$. In addition, we found a $T^4$ power law behavior in the specific heat below the ordering temperature possibly due to the gapless mode associated with either quadrupole or octupole ordering. Moreover, the specific heat exhibits the $T^{-1/2}$ dependence above the ordering temperature which is also consistent with the previous report made on the sample with RRR ~ 6 [12].

## 2. Experimental

The single crystals of $PrV_2Al_{20}$ were grown by Al self-flux method, using 4N (99.99%)-Pr, 3N-V and 5N-Al, respectively. Here, we have succeeded in getting high quality single crystal mainly by tuning the starting ratio. The RRR of the samples which were defined by $\rho$ (300 K) / $\rho$ (0.3 K) at zero magnetic field is estimated to be 20. The sample size is about 0.2 mm × 0.2 mm × 0.6 mm and the weight is 0.111 mg.

The specific heat was measured by a relaxation method. Measurements below $T$ ~ 1 K were made by using a specific heat cell installed in a $^3$He-$^4$He dilution refrigerator. For the measurements above $T$ = 0.9 K, a commercial system (PPMS, Quantum Design) was used.

## 3. Results and discussion

Figure 1 shows the temperature dependences of the specific heat ($C$) divided by temperature ($T$), $C/T$, of $PrV_2Al_{20}$ for the samples with RRR ~ 20 (filled circles), RRR ~ 7 (open circles) [23] and RRR ~ 6 (open squares) [12], respectively. As is clearly seen in the figure, the peak structure becomes sharper as RRR increases. A broad single peak observed for RRR ~ 6 separates into a sharp double peak structure for RRR ~ 7 and 20. The entropy values associated with these transitions are roughly the same among the sample with RRR ~ 20 and ~ 7 (not shown). As it was discussed for the sample with RRR ~ 20, this reaches $(1/2)R\ln 2$ at around the transition temperatures and $R\ln 2$ at a few K [20]. Following the previous work [23], we refer to these two peaks as $T_Q$ at higher temperature and $T^*$ at slightly lower temperature. The double peak structure is more pronounced in the sample with RRR ~ 20. Both $T_Q = 0.75$ K and $T^* = 0.65$ K are ~ 0.05 K higher in RRR ~ 20 compared to those in RRR ~ 7. In addition, the peak height at $T^*$ is 30% larger in RRR ~ 20 than the one in RRR ~ 7. These strongly suggest that the double peak structure is the intrinsic behavior. Note that there is no hysteresis between the data obtained on cooling and warming through the transition at zero field. The higher multipolar transition temperatures in the better quality samples have been also suggested from the previous resistivity measurements [20]. The origin of the double peak structure is unknown. Given the fact that both quadrupole and octupole degrees of freedom are available in the cubic $\Gamma_3$ doublet state, one interesting possibility is the octupole and the quadrupole orderings at $T_Q$ and $T^*$, respectively. This is an interesting future issue to be examined.

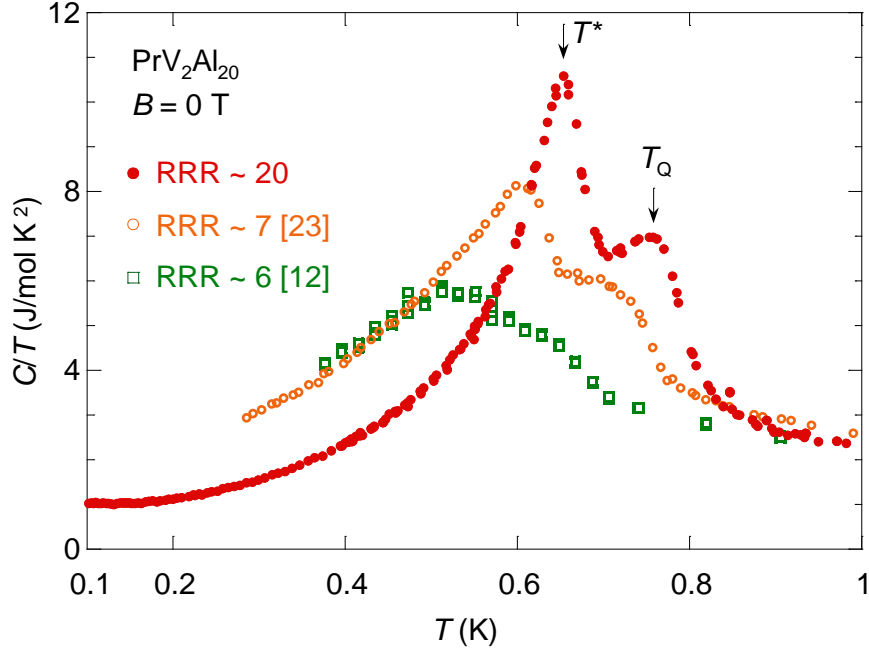

**Figure 1.** (Color online) Temperature dependence of $C/T$. Filled circle represents $C/T$ of the sample with RRR ~ 20, and open circle and open square represents the previous results obtained for the sample with RRR ~ 7 [23] and 6 [12], respectively.

In order to discuss the temperature dependence of the specific heat below $T^*$ and above $T_Q$ in detail, we present the $4f$ component of the electronic specific heat in PrV$_2$Al$_{20}$ ($C_{4f}$) versus temperature in the full logarithmic scale in figure 2. Here, $C_{4f}$ is obtained after subtracting the lattice contribution using $C$ of LaV$_2$Al$_{20}$. At the lower temperatures below 0.5 K, $C_{4f}$ is well expressed by $C_{4f} = \gamma_0 T + \alpha T^4$ as shown by the solid line in figure 2. After subtracting $\gamma_0 T$ contribution with $\gamma_0$ ~ 0.9 J/mol K$^2$, we could get the data represented as filled circle in figure 2. Normally, an excitation gap is expected below the quadrupole ordering temperature due to its anisotropic character. Indeed, PrTi$_2$Al$_{20}$ shows the exponential $T$ dependence of the specific heat and resistivity below $T_Q$ [24]. The $T^4$ power law behavior in PrV$_2$Al$_{20}$ indicates the existence of a gapless mode corresponding to the so-called "orbiton" proposed for various transition metal based systems [25]. This gapless character should be the results of the strong screening effects of the quadrupole moments by conduction electrons. The gapless mode of the quadrupole moments would be the signature of strong quadrupole fluctuations arising from the proximity to a quadrupolar quantum phase transition. In the high temperature region at $T > T_Q$, $PrV_2Al_{20}$ shows anomalous metallic behavior of $C_{4f} \propto T^{-1/2}$, consistent with the previous report on the sample with RRR ~ 6 [12]. This should arise from the strong hybridization as well.

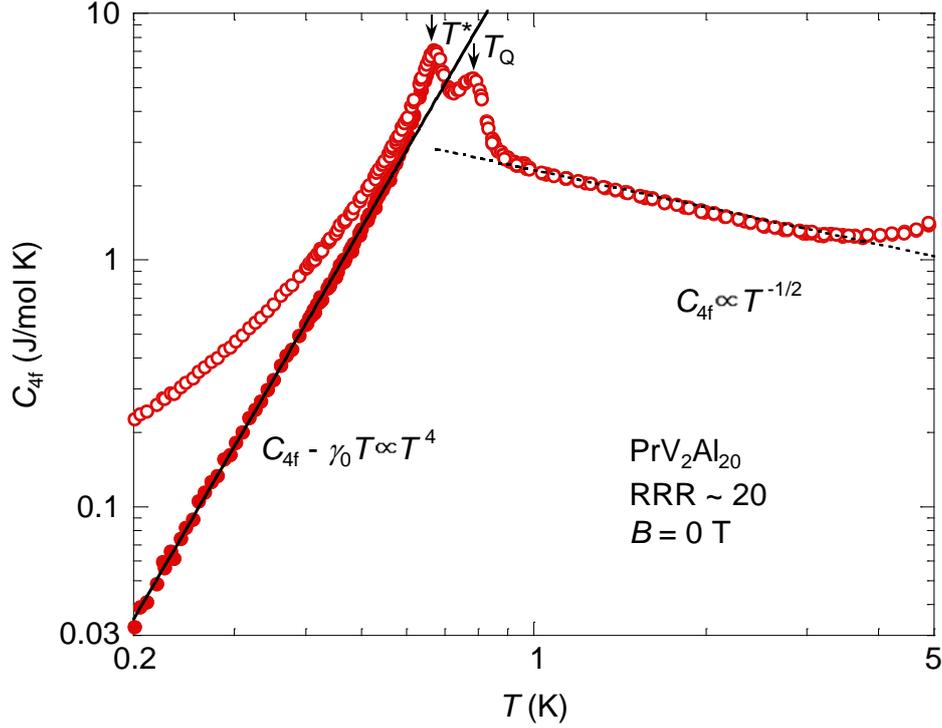

**Figure 2.** (Color online) 4$f$ component of the electronic specific heat in $PrV_2Al_{20}$ ($C_{4f}$) versus temperature in the full logarithmic scale (open circle). $C_{4f} - \gamma_0 T$ is also shown below $T^*$ in order to extract the $T^4$ power law behavior (filled circles). The solid line and dot line represent $T^4$ and $T^{-1/2}$ dependence, respectively.

## 4. Conclusion
We measured the specific heat of pure single crystal of $PrV_2Al_{20}$ with RRR ~ 20. The high quality single crystal allows us to observe the sharp double transitions at $T_Q = 0.75$ K and $T^* = 0.65$ K. The double peak structure is sharper than the one seen in the previous report [23], indicating this is the intrinsic behavior in $PrV_2Al_{20}$. In the ordered state, the $T^4$ power law dependence in the specific heat suggests the existence of a gapless mode due to strong quadrupolar fluctuations. The anomalous $T^{-1/2}$ dependence at $T > T_Q$ due to the strong hybridization was also observed in the high quality sample. One interesting possibility for the origin of the double transition is the quadrupole and octupole ordering at $T^*$ and $T_Q$, respectively. To clarify the origin, the magnetic field dependence of the double peak structure in various field directions will be important. In addition, μSR and NMR measurements using the high quality samples will be also useful for the purpose such as to detect the time-reversal broken symmetry due to the octupole ordering.

## 5. Acknowledgments

We thank Y. Uwatoko, K. Matsubayashi, J. Suzuki, T. Sakakibara, Y. Shimura, K. Araki, Y-B. Kim, K. Kuga K. Hattori and K. Ueda for supports and useful discussions. This work was partially supported by Grants-in-Aid (No. 25707030) from the Japanese Society for the Promotion of Science, PRESTO by JST, and by Grants-in-Aids for Scientific Research on Innovative Areas ``Heavy Electrons'' of the Ministry of Education, Culture, Sports, Science and Technology, Japan. The use of the facilities of the Materials Design and Characterization Laboratory at the Institute for Solid State Physics, The University of Tokyo, is gratefully acknowledged. One of the authors (M. T.) was supported by Japan Society for the Promotion of Science through Program for Leading Graduate Schools (MERIT).